\documentclass{sigchi}

\toappear{
	Submitted for review.
}

\usepackage{balance}  
\usepackage{graphics} 
\usepackage{times}    
\usepackage{url}      

\usepackage[utf8x]{inputenc}
\usepackage{ucs}
\usepackage[T1,TS1,T3]{fontenc} 
\usepackage{amsmath,amsfonts,amssymb,textcomp,ae,aecompl,aeguill} 

\usepackage[american]{babel}
\usepackage[babel]{csquotes}
\usepackage{microtype}

\usepackage{graphicx}
\usepackage[font=scriptsize]{subfig}

\usepackage{tabularx}
\usepackage{array}
\usepackage{booktabs}
\usepackage{multirow}
\usepackage[para]{threeparttable}
\usepackage[table]{xcolor}

\usepackage{listings}

\usepackage{babelbib}

\makeatletter
\def\url@leostyle{%
  \@ifundefined{selectfont}{\def\UrlFont{\sf}}{\def\UrlFont{\small\bf\ttfamily}}}
\makeatother
\def\pprw{8.5in}
\def\pprh{11in}

\newcommand\tabhead[1]{\small\textbf{#1}}

\begin{document}

\title{Appropriation as neglected practice in communities: presenting a framework to enable EUD design for CoPs}


\maketitle



\begin{abstract}

Communities present considerable challenges for the design and application of supportive information technology (IT), especially when these develop in loosely-integrated, informal and scarcely organized contexts, like it is often the case of Communities of Practice (CoP). An approach that actively supports user communities in the process of IT appropriation can help alleviate the impossibility of the members of these communities to rely on professional support, and enable even complex forms of tailoring and End-User Development (EUD).
Although this approach has been already explored by an increasing number of researchers, however there is still a lack of a general framework that could play a role in the comparison of existing proposals and in the development of new EUD solutions for CoPs. The paper proposes a conceptual framework and a related architecture, called Logic of Bricolage, that aims to be a step further in this direction to enable better EUD-oriented support for digitized communities. The framework is described and the architecture instantiated in three concrete EUD environments that specifically regard collaborative activities in order to show the generality and applicability of the framework.
\end{abstract}

\section{Introduction}
\label{sec:intro}

According to a oft-cited defnition, ``appropriation'' can be seen as ``the process by which people adopt and adapt technologies, fitting them into their working practices''~\cite{dourish_appropriation_2003}. This process has an important role in the perpetual evolution of communities, as it concerns patterns of technology adoption that can only be learnt through situated practice and social participation, as well as ``the transformation of practice at a deeper level'' than mere customization (ibidem).

Moreover, communities appropriate the technologies that mediate interactions among their members in complex and partly unanticipated ways~\cite{pipek_supporting_2006}; this is because this process is intertwined with a great deal of invisible work, tacit knowledge, conventions, habits and mutual expectations, all elements that have always been recognized as essential elements in the constitution of Communities of Practice (CoPs)~\cite{wenger_communities_1998}. In this light, technology appropriation regards the way in which technology comes to play a role within the system of meanings that is reflected in the practices shared within a community and get intertwined with the ways meanings are created and developed within a CoP~\cite{dourish_appropriation_2003}. 

As appropriation concretely relates also to specific ways to configure, adapt and tailor technologies, we share the wonder by Bodker~\cite{bodker_when_2006}, who observed how end-user tailoring is seldom taken in serious consideration when speaking of ``design for communities''. Here we are referring to tailoring not in terms of the individual adaptation of technology for personal use (which is the narrow way in which it is usually considered) but rather to the ``adaptation and further development [of computational technologies] through interaction and cooperation among people'', which calls for specific methods and environments that enable end-users to create and maintain their own tools.

The idea that computational artifacts can be developed by end-users themselves is not new: it has been the main tenet of research fields like End-User Programming (EUP) and End-User Development (EUD) since the introduction of the first computers into organizational workplaces. Yet, to date it has not yet gained in popularity in IT production, let alone in regard to CoP support. A recent and seminal analysis of the reason for this gap between research and practice asserts that such approaches ``have not been developed to cover end users' entire scope of work''~\cite{syrjanen_technology_2011}. This work is primarily social and deeply grounded in communities of practice. 

Thus, we observe a paradoxical phenomenon: on one hand, designing technological support for CoPs is seldom articulated in terms of enabling their members to autonomously build and shape their own tools, that is the main concerns of EUD research~\cite{pipek_supporting_2009}. On the other hand, EUD research seldom takes CoP as a first class concept and seems to disregard the fact that end-users are most of the times members of complex social ensembles, rather than just individual users. 

Yet, we believe that a necessary condition for EUD initiatives to keep the promise to improve technology appropriation by end-users --~and hence also increase their satisfaction (for), inclusion (into), and empowerment by the technology~-- lies in the recognition of the collective, collaborative and mostly tacit nature of technology tailoring and adoption, which underpin practices that themselves can bind a community together. Likewise, we also believe that the delicate process by which a community learns how to appropriate and exploit a technology mediating its practices and interactions can be fostered by adopting some tenets of EUD and developing them in a community oriented perspective. To this aim, previous EUD contributions do not need to be discarded, but rather to be presented in a more organic and coherent way to this ambitious goal.

To fill in this gap in literature,
the paper aims to contribute to bridge this apparent gap by submitting a purposely general framework that could help interested researchers compare existing EUD initiatives and leverage previous EUD experiences to design better CoP support. To this aim, we propose a ``Logic of Bricolage'' (LOB) as a sort of formal foundation for the conception of EUD enabling environments to help designers use concepts that are unambiguous in meaning and definite in scope, so that next researches can focus on higher level concerns, like those at application and community level.

The paper continues presenting
the instantiation of the LOB framework in three different, but relevant to our aims, EUD application domains where technologies are used by and within communities of practice: the case of the construction of document-based collaborative applications; of the integration of existing devices and software components in a collaborative setting; and of the integration of heterogeneous content in collaborative mashups, be them both event-driven and data flow oriented. These three situations, that are bound together by the collaborative nature of the work that members of specific communities of practice carry on, cover a significant amount of concrete cases where users express the requirement to be in full control of the development of their computational tools and make it a collective, incremental and often bottom up, spontaneous process. The concluding section discusses the advantages of such a conceptual framework, how it is related to other proposals and the research agenda that is necessary to make it a common reference for the design of EUD environments supporting community appropriation of technology.

\section{Toward a Logic of Bricolage for EUD systems}
\label{sec:environments}

The expression \emph{Logic of Bricolage} was used by \cite{lanzara_between_1999} to point to some general features that the ``environments'' supporting bricolage as an EUD practice should provide. We chose this expression to denote a framework by which to characterize and formalize some relevant aspects that EUD environments should exhibit to support this collaborative practice, namely both editing and execution environments; these are the application context where tools are \emph{co-defined}, and \emph{used}, respectively, that is sets of functionalities that an EUD platform provides to, on the one hand, build user-defined behaviors and structures (editing environments), and on the other hand, to interpret and execute those behaviors at run time (execution environment), while the above mentioned structures are used in the daily work of their end-users.
 
\paragraph{Constructs and Structures} -
Environments supporting bricolage have to oppose traditional approaches that provide users with sophisticated (i.e., semantically rich) modeling tools and facilitate the top-down construction of applications, e.g. by conceiving the main ``entities'' involved, their attributes, their mutual relationships, and the ``business processes'' where all these latter interact. To this aim, the Logic of Bricolage requires to provide users with a set of ``building blocks'' that they can arrange and compose together in a bottom-up fashion within a conceptually consistent environment that defines their rules of composition.  For this basic concept we borrow from~\cite{lanzara_between_1999} the term \emph{construct} to denote all the basic building blocks that end-users engaged in the active task of ``bricolaging'' their tools (who for this reason in LOB are called \emph{bricolants}) can identify - and possibly construct - as the atomic entities that the editing environment should make available; the term \emph{structures} is used to denote the working spaces and artifacts, as well as their computational behaviors, that are constructed and used by the bricolants by means of the constructs  previously identified.  

We further distinguish between \emph{Operand Constructs} and \emph{Operator Constructs}: operands are the most atomic data structures, components and variables that make sense in that domain (see Production 4 in Figure~\ref{fig:formalization}); operators are all the feasible operations and micro-functions that users deem necessary to be performed over the operands in their application domain; operators can be either \emph{functional} or \emph{actional} (see Production 5), to indicate something either akin to the evaluation of a function, or related to the production of some actual effect in the computational environment, respectively; in particular, (functional) Operator constructs can be applied to operands to allow for the recursive construction of more complex operands from simpler ones. 

According to the bottom-up approach advocated within the LOB framework, these both kinds of construct have to be identified during the inception phase of the platform within a cooperative setting or organization, as a result of the collaboration among the members of a CoP and the IT professionals. In fact, not every kind of construct is easy to build: for example, according to the specific application domain (e.g., information systems, computer-assisted design), users and developers can decide to implement operator constructs by specializing and articulating a sequence of primitives that the platform exposes natively (see more below). In doing so, the work of ``constructing'' the basic operations  that it would more convenient to consider as ``atomic'' (i.e., black box) later by users is facilitated. 

\begin{figure}[tb]
\begin{lstlisting}[basicstyle=\scriptsize,breaklines=true,breakatwhitespace=true,frame=shadowbox]
Generative Productions of the Logical Of Bricolage
---
1) <web-structure> ::=  <layout-structure> +
2) <layout-structure> ::= <topological-object>+
3) <topological-object> ::= <operand-construct> <coordinates>?
4) <operand-construct> ::=  <constant> | <typed-variable> | <operator-construct>(<operand-construct>+) 
5) <operator-construct> ::= <functional-operator-construct> | <actional-operator-construct> 
6) <annotation> ::= <style> <target-ref>+ | <constant> <target-ref>+
7) <target-ref> ::=  <functional-operator-construct> (<target>)
8) <constant> ::= <domain-values> | <multimedia-text> 
9) <target> ::= <control-structure> | <topological-object> | <annotation>
10) <style> ::= <conventional-symbol>+ | operator-construct(<target>)
11) <control-structure> ::=  <rewriting-rule> |   <connector>  
12) <connector> ::  <functional-operator-construct> (<rewriting-rule>+)
13) <rewriting-rule> ::=  <condition>* <action>+
14) <condition> ::= <functional-operator-construct> (<state>)
15) <action> ::=  <actional-operator-construct>(<state>)
16) <state> ::= <operand-construct>+
---
Legenda: The LOB grammar is expressed in EBNF-like notation; therefore, the symbol `|'  means  `alternative'  ; `<>' means  'variable'; `+'  means `one or more occurrences';  `*' means 'zero or more occurrences'; `?' means `zero or one occurrences'; ``domain values'' are not specified and are the terminal symbols of the Grammar (e.g. True and False).
\end{lstlisting}
\caption{A preliminary formalization of the Logic of Bricolage}
\label{fig:formalization}
\end{figure}

\paragraph{Editing and Execution Environments} -
Arranging constructs into suitable structures requires an editing environment by which users can shape the data structures and related logic that are needed for the desired application. We distinguish between \emph{Layout structures} and \emph{Control structures}\footnote{We prefer the expression ``layout structure'' instead of ``information structure'' (or ``data structure''), as we mean to hint at the material, spatial arrangement of meaningful signs that ``act at the surface'' in promoting cognitive processes of sense making and interpretation.}. The former ones are sort of material (yet non necessarily tangible) and symbolic ``work spaces'' that are recognized by members of a community of users as the physically inscribed and computationally augmented artifacts
where and by which they carry out their work. In the domain of computer-aided design and collaborative drawing/editing, a Layout Structure is the working space where users arrange the docking bars of their preferred commands, symbol stencils and predefined configurations of elements that must be set up before the actual work begins. In document-based information systems, Layout Structures are the document templates of forms and charts that are used to both accumulate content and coordinate activities
; such structures are endowed of both physical properties and symbolic properties, for instance, input controls (i.e., data fields), boilerplate text, iconic elements and any visual affordances conveyed through the graphical interface (see Production 9 and 10), which in LOB are all instances of Operand constructs (see Production 4). Accordingly, Productions 2 and 3 express the fact that Layout structures result from the topological arrangement of Operand constructs. With reference to Production 1, Layout structures can be aggregated in Web Structures, that are recursively defined as interconnected sets of Layout Structures, and hence topologically unique arrangements of Operand Constructs (see Production 1, 2 3 and 4).  

Also Control Structures are recursively defined in terms of simpler rewriting rules (see Production 11), that is sort of ``conditioned actions'' (Productions 13), which are expressed in terms of specific Operator Constructs (see Productions 14 and 15). 
Conditions are expressed as functional constructs applied to the current state of computation: more precisely, the state encompasses all application data, and the functional constructs are Boolean functions, whose union set is considered ``true'' according to the current state, if and only if each of these functions returns true on its input values.
Control Structures can be of arbitrary complexity, from simple rules to sequences of instructions in virtue of Connectors (see Productions 11 and 12), which are particular functions that make Control structures recursive (for instance, connectors can be parametric NAND and NOR functions, by which to compose any other logical sequence of high-level instructions). In simple terms, \emph{Control structures} specify the behaviors of Layout structures, i.e., how the artifact acts on the content inscribed therein, e.g., in response to events generated at interface level, and how this level interacts with users during the habitual use of the application. This kind of structure is interpreted (and hence executed) by the execution environment: as said above this is the running application that has been previously constructed in the editing environment by end-users and designers: such an environment executes Operator Constructs (both functional and actional ones) by interpreting them as more or less complex articulation of Primitives; these latter, in their turn, are domain-independent functionalities exposed by the platform that have been expressed in terms of lower level Application Programming Interfaces by IT professionals. It is noteworthy the fact that Operator constructs can be just domain-specific specializations of primitives, like, e.g., a user-defined procedure \textit{sum()} can specialize the \textit{+} primitive of a programming language.

\paragraph{Annotations} - We call annotation any user-defined content that is created to be anchored to another content and conceived to \emph{supplement} such a content
. Also annotations are a first-class concept of the LOB framework, for their central role in collaborative work in supporting work articulation, knowledge sharing and mutual understanding~\cite{luff_tasks--interaction:_1992,cadiz_using_2000,bringay_annotations:_2006,ardito_end_2012,cabitza_leveraging_2013}.

For their supplementary nature, annotations act at a more informal level than institutionalized structures and the official content that is accumulated therein during situated practice, and as such they cannot be fully predermined at definition time; rather, the platform must expose specific primitives for their addition, deletion, etc. within the execution environment, that is when users actually use their application. 
Annotations can be either stigmergic signs and marks attached to the content of a document or any ex-tempore comment and semantic tags chosen by users from either domain specic taxonomies or setting-specific
folksonomies (see Productions 6 and 7 in Figure 1). The first case is expressed by <style>, which can be either a symbol or an operator construct that is applied to the target content to change its format or visual rendering (see Production 9); the second case is expressed in terms of a <constant>, which can be any value or multimedia text (see Production 8). Notably, in LOB annotations can be nested (see Production 9, where targets can be annotations as well), that is users should be able to \emph{annotate annotations}, so as to allow for nested threads of comments and tags, as we described 
in [anonymized]. These written texts (i.e., constants in LOB, see Production 8) are all conceived as pieces of a collaborative and never-really-finished bricolage, which hosts informal communication and handover between practitioners, their silent and ungoverned work of meaning reconciliation, and the sedimentations of habits and customs in effective (yet still unsupported computationally) conventions of cooperative work~\cite{cabitza_leveraging_2009}. 

\begin{figure*}[tbh]
  \centering
      \includegraphics[width=.85\textwidth]{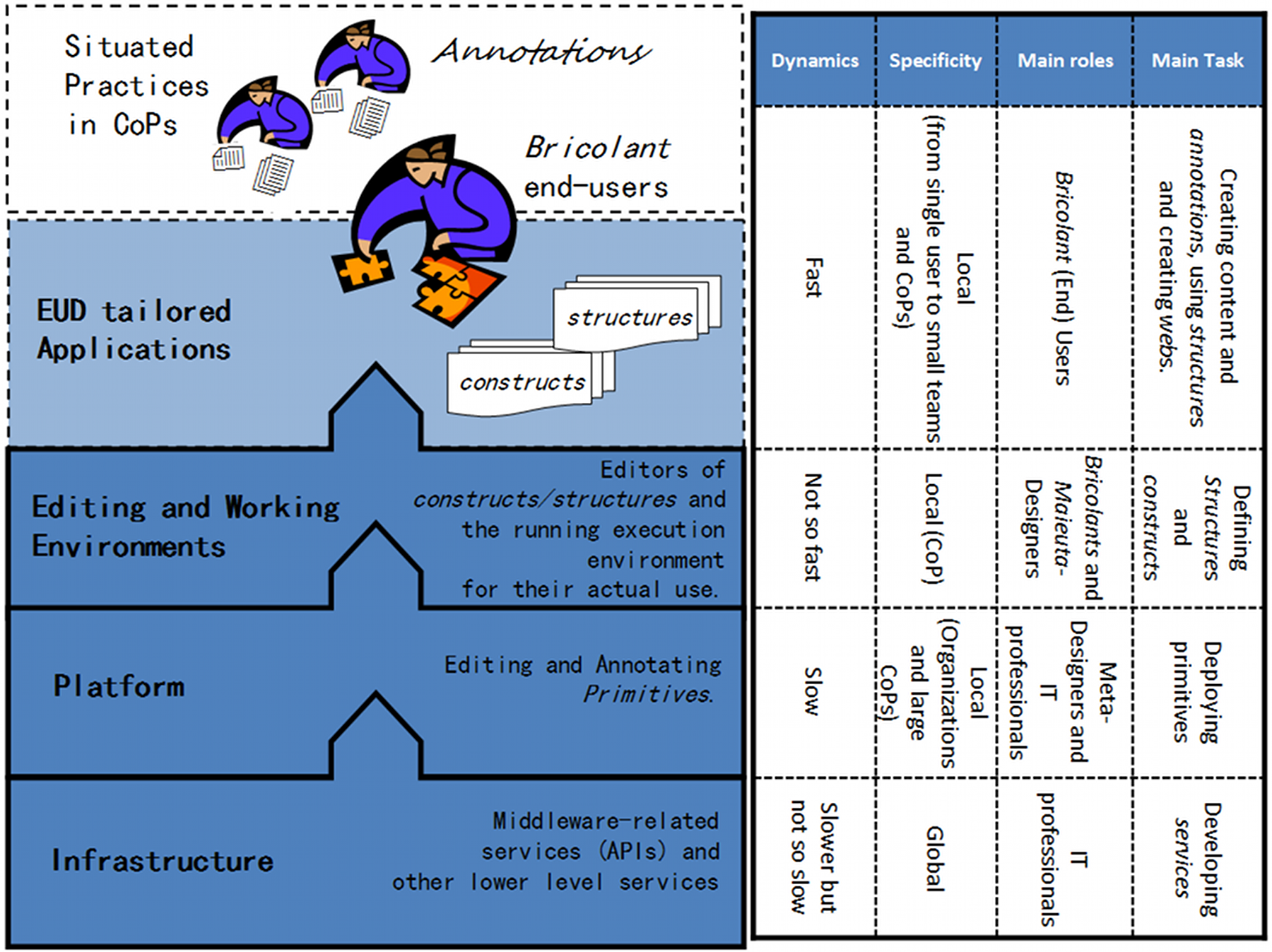}
  \caption{A conceptual architecture for an environment supporting EUD bricolage. LOB keywords are in italics; for each layer, its name and what it offers to the higher layers are specified.}
  \label{fig:architecture}
\end{figure*}

All the LOB elements characterized in this section can also be arranged into a conceptual architecture, that we depict in Figure~\ref{fig:architecture}. With reference to this layered architecture, it is noteworthy to stress that end-users are enabled to create community-specific \emph{applications} by interacting with \emph{environments} that, to this aim, expose apt building blocks (i.e., constructs) through \emph{services} (e.g., specific editors) that are supplied by the underlying EUD \emph{platform}; this latter one in its turn is enabled by a regular infrastructure (i.e., an application server and operating system). Applications are tailored and appropriated through incremental and actually never-ending task-artifact cycles~\cite{carroll_task-artifact_1991} in which members of a CoP agree over time upon what constraints and functionalities must be enacted by their supporting technology.

\paragraph{Between transiency and permanency} - 
In complex systems (as socio-technical systems are) change must be expected: this has become a sort of common place in IT system design since its foundation as a scientific discipline in the 1960s; we can relate it to the vast research undergone about exception handling in software engineering so far and in what has been called the ``struggle for flexibility'' 
; yet the effect of change must be recognized as intrinsically unpredictable, and this is a fact that is less recognized than it should, at least in practice. Thus, in~\cite{mansfield_nature_2010} Mansfield argues that `if the majority of computer-based socio-technical systems fail to meet the expectations of their sponsors, perhaps that is due to their [too rigid] architecture''; consequently he submits design-oriented principles that call, among others, for the adoption of a layered architecture where some layers are allowed to change (and adapt to changes, or evolve in response to them) at \emph{different rates} to account for the varying rate of impact of the events affecting them. The LOB conceptual architecture that is depicted in Figure~\ref{fig:architecture} recognizes this point, as it encompasses different layers that account for both different scopes (see Specificity), concerns (see Task), involved roles and dynamics. 

We borrow again some terms from  Lanzara's Logic of Bricolage to qualify the changing rate of the layers constituting  the conceptual architecture depicted in Figure~\ref{fig:architecture}. 
Any form of annotation carried out by practitioners \emph{over} and \emph{upon} structures and their content can be seen as a \emph{more transient}, informal and more user-driven piece of bricolage, which acts at a sort of different layer with respect to content, structures, constructs and obviously the platform's  primitives (see  Figure~\ref{fig:architecture}); nevertheless (or right in virtue of this complementarity), annotations play an equally important role in making the artifacts-in-use flexible enough to support also invisible work and hence fully appropriated by their users. Layout structures are the immediately less transient components as they correspond to working spaces that users flexibly accomodate to their changing needs\cite{harris_better_1999}. Decoupling layouts, i.e., data structures, from the logic acting upon them, i.e., control structures, is a well known engineering principle that the framework recognizes. Yet, although formally decoupled Control structures in practice follow and support the modifications of the objects/artifacts to which they are to apply: therefore, they are at the same level of change rate. On the other hand, constructs, especially the more basic and atomic ones, can be considered as changing at a slower pace than the structures they are part of, while the more complex ones can be revised (modified, deleted) more often but surely less frequently than the above mentioned structures. Primitives, conversely, are \emph{quasi persistent}: indeed, the composition rules must be more stable than the objects they apply to. Finally, the layers of the underlying technological infrastructure can be considered as almost \emph{persistent}, in that changes can be planned, postponed, made incrementally depending on the triggering technological evolution or organizational strategy. The emphasis on this aspect is motivated by the fact that IT professionals have to build the infrastructure and its relationships with constructs and primitives plastic enough to avoid any friction between the different layers that ``drift'' at different speed according to regular technological evolution and the users' needs.

In the next section we illustrate how the LOB conceptual architecture can be instantiated to deal with concrete, although quite general, EUD practices to support real work settings. To this aim, the formalization of the provided in Figure~\ref{fig:formalization} will help identifying the components to be instantiated and to be offered to bricolant end-users through the editing environments.


\section{Instantiating LOB for specific contexts of use}
\label{sec:instances}

This section describes three concrete environments that we reinterpret as instantiations of the LOB conceptual architecture: the first environment is a document-based collaborative system tailorable to the local work setting; the second environment allows for the user defined integration of devices and software components supporting  groups of cooperating actors; the third one is an example of a multi-layered and flexible mashup composition environment that allows for the integration of data sources and functional components to produce enriched and personalized results.

\subsection{WOAD: constructing Webs of Active Documents}
\label{sec:woad}

In the recent years the field studies we conducted in different work settings, especially in the healthcare domain, led us to identify the functionalities of a document-based collaborative system, called Web of Active Documents (WOAD)~\cite{cabitza_woad:_2010,fred_web_2011} that are currently under incremental development. 
The core concepts of WOAD can be summarized as follows in terms of: i) the information structure is composed of hyperlinked active documents that can be annotated in every parts and sections and be associated with any other document, comment and computational behavior; ii) there is no rational and unified data model: users define their forms in a bottom up manner and, in so doing, the platform instantiates the underlying flat data structures that are necessary to store the content these forms will contain and to retrieve the full history of the process of filling in them; iii) the presentation layer is in full control of end users, who are called to both generate their own templates and specify how their appearance should change later in use under particular conditions by means mechanisms that are expressed in terms of if-then rules. Users can define local rules that act on the documents' content and, as hinted above, change how documents look like (i.e., their physical affordances), to make themselves aware of pertinent conditions according to some cooperative convention or business rule like, e.g., the need to revise the content of a form, or to consider it provisional, or to carefully consider some contextual condition 
. The LOB conceptual architecture offers a framework that incorporates the various WOAD's components in a coherent picture  as follows.  In the following we associate the concrete items constituting WOAD with items of the LOB framework. 

First, we specialize the constructs: remember that these are application domain dependent and therefore they have to be defined by the users in cooperation with designers and, when necessary, with IT professionals according to the needs of the specific application domain. 


The \textbf{Operand constructs} in WOAD are called datoms (document atom): these are but any writable area with a unique name and a type (e.g., Integer, String).  A datom can recursively be a composition of one or more datoms:  e.g., the `first name' datom (a string) and the `family name' one (also a string) can be combined into a `person name' datom that encompasses both. 
The \textbf{Operator constructs} are a selection of atomic functions (including predicates, denoted as functional in Figure~\ref{fig:formalization}): for example, doctors from a medical setting described in~\cite{cabitza_leveraging_2009,cabitza_leveraging_2013} required, besides the standard arithmetic and boolean operations, also a construct to perform the \emph{average}, and another one checking the occurrence of a value in a given set (the \emph{is-in} construct). A list of Actional Operator Constructs conceived to be applied to the Layout Structures or on their components has been derived from a series of field studies and it encompasses basic operations like \emph{save}, \emph{retrieve} and \emph{store} (see \cite{cabitza_remain_2011} for the full list); one of these constructs, namely \emph{annotate}, can associate a didget (i.e., a document widget resulting from the instantiation of a datom) with an annotation.
More complex operator constructs can be recursively defined by composing more elemental ones. 
\textbf{Web structures} are a graph of hyperlinked \emph{templates}, i.e., WOAD \textbf{Layout Structures}; these latter are a set of didgets,: a didget is a topological object, i.e., an Operand construct called datom (see above) that is put in some place, i.e., is coupled to a set of coordinates (that in WOAD are represented as Cartesian pairs with respect to the origin of the template). It is worthy of note the fact that the two datoms mentioned above (first name, family name) can be used to create a WOAD template (i.e., a Layout Structure), as well as a third datom, i.e., another Operand construct: in the former case the two-datom set is to be used in the execution environment in documentts that are instances of the template encompassing it; in the latter case the set is intended to be used atomicaly as component itself of other templates (i.e., as a topological object) in the editing environment.
In WOAD the \textbf{Control structures} are called \emph{mechanisms}, i.e., if-then rules whose if-part is a Boolean expression that is recursively defined using the predefined datom names as variables, and the operators identified above, all together with the (obvious) constants of the basic types.  The then-part is a sequence of actions that has to be performed on the template or on its inner components. Mechanisms are connected by the (implicit) OR connector.
In WOAD, the \textbf{Primitives} that allow for the definition of both Layout and Control Structures are the following ones: \emph{aggregate}, to build complex operands from simpler ones; \emph{compose}, to build complex operators in terms of functional composition; \emph{localize}, to associate a didget with a Cartesian coordinate with respect to the origin of a given template; \emph{list}, to build sequences of if-then rules. Moreover, the \emph{annotate} primitive associates a text with an operand.construct, that is creates one of the pairs of values reported in Figure~\ref{fig:formalization}.


As mentioned in the previous section, the Primitives are offered through an editing environment where Constructs and Structures can be defined. In WOAD this environment is constituted by two visual editors: one for the construction of mechanisms~\cite{cabitza_providing_2012}, and one for the construction of datoms and, by arranging these latter topologically in terms of didgets, templates~\cite{cabitza_tailorable_2011}. The editing environment is associated with an execution environment that has been tested on realistic examples taken from the  healthcare domain~\cite{cabitza_woad_2011,cabitza_rule-based_2012}. 
 
\subsection{CASMAS: creating hybrid communities for cooperation.}
\label{sec:casmas}

Suppose that a set of applications and devices have to be integrated to support a set of actors that cooperate by means of them.  According to the Community-Aware Multi Agent Systems (CASMAS) framework \cite{locatelli_community_2010,locatelli_end-users_2012}, both actors and their tools all are represented as \emph{entities} and integration can be seen alike to becoming members of the same \emph{community}: as such, they coordinate their behaviors through a shared information space 
that contains coordinative information, as well as the \emph{behaviors} that are dynamically assigned to each entity to make it an active member of the community: 
in CASMAS communication is asynchronous but it is not message based. Instead, when an entity \emph{posts} a request into the space, other entities will \emph{react} to this request according to their current behaviors.

The CASMAS framework encompasses a language to specify entities and their behaviors. This language takes the declarative form of facts and rules (if-then patterns), which offers  the possibility to express behaviors in a modular way, without the need to define complex and exhaustive control structures~\cite{myers_natural_2004,locatelli_integration_2011}. The rules constituting an entity's behavior express \emph{what} the entity is expected to do \emph{when} some conditions are satisfied: these conditions are matched against the facts contained in the community's space and in the entity's local memory;
the action(s) that the entity should perform updates either the community space or the memory of the entity itself.

The integration of a software application/device is realized by inserting a fact in the memory of the entity representing it and by defining the behavior of this entity. The fact contains attribute-value pairs that specify the information the application/device makes available for sake of coordination with the other entities of the same community; 
the entity's behavior expresses conditions (among others) on the concrete application/device attributes (when) and invokes  some of the functions the application/device exposes to the community (what): actually, the entity is a sort of wrapper that mediates between the concrete application/device and the integration environment (community).

As done for the WOAD framework, we associate each CASMAS feature with each item of the LOB framework with reference to the formalization depicted in Figure~\ref{fig:formalization}.

In CASMAS, the \textbf{Operand constructs} are the facts that are contained in and exchanged across community space(s): CASMAS facts are expressed according to the syntax of the underlying rule-based language (currently, JBoss Drools\footnote{\url{http://www.jboss.org/drools/}}.). 
The \textbf{Operator constructs} are the basic functions and predicates that are exposed by the underlying rule-based programming language;in particular the actional operator constructs (actions in CASMAS) support the asynchronous communication between entities  as well as the storing and retrieval of information among the spaces and the local memories. In case of applications/devices' memories, the store/retrieve actions respectively put and get information to/from the data structures therein managed.
The \emph{spaces} that are implicitly connected through the entities that are members of more than one community are the LOB \textbf{Web structures}; each space is a \textbf{Layout structure} that contains the community's facts and the entities' ones; differently from WOAD, facts (i.e., operand constructs) are not geometrically localized within spaces, as CASMAS does not specify the coordinates of its topological objects (i.e. the facts within the spaces).
The if-then rules connected through the OR connector and grouped according to the entities membership to the community are the CASMAS \textbf{Control structures}: their if-parts encompass sets of Operand and Operator constructs, as in WOAD; similarly, the then-part can either encompass the above mentioned \emph{put} and \emph{get} actional operator constructs, whenever the behavior regards applications/devices entities; or a \emph{post} action, in the other cases. The CASMAS framework defines the same primitives seen for WOAD (except for the \emph{localize} and \emph{annotate} primitives), but it also encompasses the \emph{put} and \emph{get} primitives: the role of these primitives is to interact with the wrappers developed for each devices/application to be integrated and they are called in the actions having the same name.

\subsection{DashMash: flexible configuration of EUD mashups}
\label{sec:cmashup}

Recently, an increasing number of environments where users can combine information flows from different data sources, the so called mashups, has been proposed, also for commercial use (e.g., Yahoo Pipes). For sake of exemplification, we apply our exercise of LOB instantiation to the the DashMash framework~\cite{cappiello_enabling_2011}\footnote{This task is less detailed than in the other two cases as the mentioned paper does not give all the necessary details.}, which we take as representative of a wide class of applications that allow for the collaborative user-driven aggregation of heterogeneous content. Indeed, \textit{DashMash} is a general-purpose EUD environment that adopts an approach in which the design-time and the use-time are strictly intertwined: end-users can autonomously define their own mashups and execute these latter ``on the fly'', to progressively check the result of their editing activities. Like most of the traditional approaches for the creation of mashups, also the DashMash approach is dataflow-oriented, i.e., end-users can only aggregate, filter and display data in the most meaningful way, e.g., a pie chart, a table or a map. On the other hand, the DashMash approach gives also the possibility to provide end-users with an environment that can be customized so as to meet their domain-specific requirements; 
essentially, this can be done in two ways: (i) through the development of domain-specific components that allow to interact with the functionalities provided by any kind of (local or remote) service; and (ii) providing end-users with the access to data coming from private and domain-specific data sources, in addition to publicly accessible ones. Nevertheless, the approach used in DashMash provides end-users with an abstraction that makes them able to use the various mashup components (e.g., data sources, filters and data viewers) that are automatically composted on the basis of a pre-defined set of compatibility constraints, relieving end-users from the need to know any technical detail about the used components.

As for CASMAS, the DashMash Control structures are grouped to form the behavior of each \emph{component}. These constructs allow for typical publish and subscribe patterns, like ``if a new fact occurs, then publish an event'' and ``if a subscribed event occurs, then perform some operations''
New facts or  operations pertain to single components only. For example, if the component is the \emph{Composition Handler}, then the new fact is any change in a component; the components influenced by this kind of event (i.e., the subscribers) activate the corresponding operations: for example, if the change is about a Filter Component then all Data Components using this filter activate internal operations to send the data to the new Filter Component and at the end this later notifies that new-data are ready: this event is consumed by the Viewer Component subscribing this event for the specific data. In this view, the Operand constructs are the \emph{data} and the \emph{events}, while the main Operators constructs regard the publication of an event, and the subscription for a specific kind of event.  In DashMash, the Web structure is the set of workspaces; each workspace is a Layout structure that is composed by two inner Layout structures: one contains the output of all the Viewer Components for what concerns the data (i.e., at the use level according to~\cite{ardito_end_2012}); the second contains a standard description of the workspace \emph{state} in terms of Components  such as: Data Sources, Filters and Viewers (i.e., at the design level).

More traditional mashups that, differently from DashMash, are uniquely based on data flows can be described as  graphs whose nodes are input-output transformations, and whose arcs express the kind of connection that hold between two nodes. In LOB terms, a mashup belonging to this class can be seen as a set of rewriting rules that transform inputs into outputs, where arcs are as connectors that express the appropriate structure of the data flow (e.g., either alternative or parallel flows). 

Table~\ref{tbl:summary} highlights how the LOB approach applies to the the three frameworks characterized above. These three instantiations support our claim that the LOB architecture is at the same time general enough to formally describe different types of EUD application classes (e.g., information mashup, document-based systems, integration of applications), and yet detailed enough to define a concrete platform to apply recurring design patterns for EUD systems to be deployed in different application domains.

\begin{table*}[tb]
\centering
  \begin{tabularx}{\textwidth}{|l|p{2.5cm}|X|p{2.5cm}|l|}
    \hline
    \tabhead{Framework} & \tabhead{Primitives} & \tabhead{Constructs} & \tabhead{Structures} & \tabhead{Annotations} \\
    \hline
    WOAD &
    \textit{\texttt{aggregate}, \texttt{annotate}, \texttt{compose}, \texttt{list}} and \textit{\texttt{localize}} &
    \textit{\texttt{annotate}, \texttt{attach}, \texttt{average}, \texttt{cache}, \texttt{copy}, \texttt{correct}, \texttt{count}, \texttt{create}, \texttt{datom}, \texttt{delete}, \texttt{is-in}, \texttt{officialize}, \texttt{open/read}, \texttt{print}, \texttt{protect}, \texttt{retrieve}, \texttt{save}, \texttt{select}, \texttt{store}, \texttt{transmit}} and \textit{\texttt{write}} &
    \textit{\texttt{mechanisms}} and \textit{\texttt{templates}} &
    Yes \\
    \hline
    CASMAS &
    \textit{\texttt{aggregate}, \texttt{compose}, \texttt{get}, \texttt{list} and \texttt{put}} &
    \textit{\texttt{get}, \texttt{post}, \texttt{put}}, rule patterns and facts &
    \textit{\texttt{space}}\,and \textit{\texttt{behaviors}} &
    No \\
    \hline
    DashMash &
    data and events &
    \textit{\texttt{publish}, \texttt{subscribe}}, and components, data and events &
    \textit{\texttt{workspaces}} and sets of\,\textit{\texttt{workspaces}} &
    No\\
    \hline
  \end{tabularx}
\caption{Synoptic table of how LOB concepts can be applied to the frameworks under analysis.}
\label{tbl:summary}
\end{table*}



\section{Discussion}
\label{sec:discuss}

We are convinced that in EUD ``the best is yet to come'', and that the key factor for this to happen is to ``scale up'' the experiences collected in the last ten years or so of research in EUD and focus the next proposals in this strand in terms of support for the community appropriation of its technologies.

This is not to discard what has been done so far in this research field, but rather to recognize that the  solutions that have been brought forth to allow end-users to create and maintain their computational tools autonomously have now reached a maturity level that requires a sort of backward reflection, as well as an effort to generalize local solutions and intuitions into general insights and concepts for future reuse and discussion, especially for their application in real communities. 

To this latter regard, we have argued that in the context of CoPs EUD-oriented capabilities can represent a major turning point that could unleash (and exploit) the potential of those social ensembles to self-regulate and self-tune their practices of value creation according to situated elements that are difficult to externalize in terms of explicit requirements, and impossible to underpin in a specific deployment of a one-size-fits-all technology: in other words, the technology supporting a CoP must be appropriated by its members to really support them. In this regard, we see appropriation as the learning process where each member of a community understands what a technology can do for herself, and how it enables, constrains and shapes the community's practices, either further on from or far beyond the intentions of the technology designers.

To facilitate the ``scale up'' of promising EUD proposals to the community level, i.e., to the level where the designers of an intended support of complex ad situated collaborative activities can not be easily left in the loop of the technology evolution following its initial inception, we have conceived the general framework presented in this paper, called Logic of Bricolage. 

After~\cite{halverson_resources_2008}, who aptly provides a lens through which to consider the utility of conceptual proposals, we propose the LOB framework to: i) facilitate researchers in making sense of and describing their and others' solutions (descriptive power) to leverage lessons learned and share best practices and effective solutions; ii) help designers talk about their solutions by providing them with a common vocabulary (rhetorical power), i.e., a very concise lexicon whose available terms cover few but essential aspects of recurring EUD conceptual structures and underlying models and are defined with some degree of unambiguous formalization (see Figure~\ref{fig:formalization}). Lastly, iii) LOB is proposed to both inform and guide the design of EUD proposals that could meet the challenging requirement to let the members of a CoP develop and maintain practices of technology local tailoring and adaptation to their emerging and ever-changing needs (applicative power); or at least to foster discussion on the need of such a framework in the hybrid field where CoP-oriented and EUD-related concerns meet, and the dissemination of such concerns in multiple venues, research initiatives and digitization projects. 

Differences and complementarities of LOB with respect to existing frameworks are to be found in all these three dimensions. We are aware that LOB shares some strong affinities with the approach described in~\cite{costabile_visual_2007}, which also acknowledges the substantial continuity between design and usage of software applications. The concept of Software Shaping Workshops (SSWs) therein described also allows for the definition of environments that are shaped according to specific user communities where users can use or tailor the software tools supporting their working practices. 
Yet, unlike our proposal, the SSWs approach is more aimed at the definition of an organization structure and of a methodology for EUD design, rather than at the definition of a conceptual framework and architecture supporting the description and shaping of each possible EUD environment; even more importantly, such an approach is strongly oriented at the shaping of the interaction of users with their tools (which are however constructed by IT professionals), rather than at informing the users' activity of autonomously defining their tools themselves. This is important especially in a community-oriented perspective since, perhaps differently from structured corporate and organizational settings, CoPs evolve in mostly unanticipated ways, and are sort of ``autonomous'' bodies with respect to top-down ordering initiatives, within the scope of their constituting practices and artifact use.


\section{Concluding remarks}
\label{sec:concl}

In this paper we have presented the LOB framework, in order to provide designers a tool to identify and separate concerns in the conception and development of EUD-enabling platforms, associate these concerns with specific layers of a common reference architecture, and call objects pertaining to each layer with specific and evocative names, by following the precept to ``keep it simple, but not simpler''. 

In presenting this design-oriented framework, we purposely avoided to characterize who end-users are or to borrow and adapt any complex model of user roles, which would be over-ambitious (or just plain wrong) in the domain of CoPs: LOB is a (meta-)design-oriented conceptual framework where users are but active workers within a social context: what we call \emph{bricolant}, to stress the idea that they perform bricolage as they work, and that their work is always a form of ad-hoc, highly situated bricolage, which is ultimately learnt in virtue of an active participation in collective practices of sense-making and technology use. Bricolants are endowed with the only, and often irremediably tacit, expertise to make their tool really meaningful and useful in a ``system of meanings'' mirrored in the practices shared within a community. 

In the LOB light, an EUD platform supporting a community of end-users is seen as the enabling environment that allow them to make their useful tools also usable, and computationally powerful of course. To this aim, we envision a specific role, the \emph{maieuta}-designer (see
[anonymized] for a detailed characterization of this particular \emph{meta}-designer), to help bricolant users ``help themselves'' and reach an increasing level of autonomy with respect to traditional IT professionals (i.e., designers, architects and programmers), in a typical arrangement where each expert member helps the less expert ones in using the available tools proficiently.

The LOB framework is therefore intended as one step toward a shared systematization of technological approaches (both of the past ones and of those still to come) that could soon reach enough simplicity and generality (as desirable characteristics for good design and academic research) by progressively abstracting and formalizing the lessons learned and best solutions that the field has so far proposed and discussed. As such, in this paper we outlined the exercise to frame some of the existing EUD platforms specifically supporting collaborative activities in terms of LOB concepts; in so doing, we aimed to highlight its descriptive power and show its applicability in the EUD arena to make existing solutions more understandable and comparable, and the next ones stand more firmly on the shoulders of he formers.

\balance

\bibliographystyle{acm-sigchi}
\bibliography{IS-EUD2013}

\end{document}